\documentclass[11pt,reqno]{amsart}

\usepackage{amsmath,amsthm,amssymb}
\newcommand{\llbracket}{[\![}
\newcommand{\rrbracket}{]\!]}
\usepackage{mathtools}
\usepackage{hyperref}
\hypersetup{colorlinks=true,linkcolor=black,citecolor=black,urlcolor=black}
\usepackage{microtype}
\usepackage{proof}
\usepackage{tikz}
\usetikzlibrary{positioning}
\tikzset{
  redtree/.style={every node/.style={font=\small}, node distance=4.5mm and 0mm},
  rulelab/.style={midway, right=2.5mm, font=\scriptsize}
}

\newtheorem{theorem}{Theorem}[section]
\newtheorem{lemma}[theorem]{Lemma}
\newtheorem{corollary}[theorem]{Corollary}
\newtheorem{proposition}[theorem]{Proposition}
\theoremstyle{definition}
\newtheorem{definition}[theorem]{Definition}
\newtheorem{example}[theorem]{Example}
\theoremstyle{remark}
\newtheorem{remark}[theorem]{Remark}

\newcommand{\base}[1]{\mathcal{#1}}
\newcommand{\Bas}{\base{B}}
\newcommand{\BasC}{\mathcal{C}}

\newcommand\supp{\Vdash}
\newcommand{\suppB}[1]{\supp_{\!#1}}
\newcommand{\IPL}{\textsf{IPL}}
\newcommand{\hH}{\textsf{hH}}
\newcommand{\Horn}{\textsf{H}}

\newcommand{\baseext}{\sqsupseteq}
\newcommand{\enc}[1]{\llbracket#1\rrbracket}
\newcommand{\opder}{\vdash_{O}}
\newcommand{\opq}{\mathrel{?-}_{O}}
\newcommand{\At}{\mathrm{At}}
\renewcommand{\emptyset}{\varnothing}

\title{Towards an Operational Proof-theoretic Semantics}

\author{Alexander V.\ Gheorghiu}
\address{University College London}
\email{a.v.gheorghiu@ucl.ac.uk}


\begin{document}

\begin{abstract}
Model-theoretic semantics comes with a computational counterpart: to check
validity is to search for a countermodel, and tableaux, resolution, and
satisfiability solvers make the semantics something one can run.
Proof-theoretic semantics --- the younger paradigm in which meaning is given by
inferential role rather than by denotation --- has no established counterpart.
This paper proposes one: logic programming. We show that Sandqvist's
base-extension semantics for intuitionistic propositional logic is executable in
a precise sense. Bases are logic programs, and to each formula $\varphi$ we
assign a goal $\enc{\varphi}$ of a second-order hereditary Harrop language such
that $\varphi$ is supported in a base $\Bas$ exactly when $\enc{\varphi}$ is
operationally derivable from $\Bas$ read as a program. The encoding is the
Russell--Prawitz translation of the connectives with quantification restricted
to atoms, and this restriction is not a limitation but the point: it is what the
support clauses themselves demand. The correspondence is exact base by base, not
merely at the level of validity, so that intuitionistic validity becomes the
success of a single query against the empty program. Logic programming stands to
proof-theoretic semantics as tableaux and satisfiability solvers stand to
model-theoretic semantics.
\end{abstract}

\maketitle

\section{Introduction}
\label{sec:intro}

Tableaux and satisfiability (SAT) solvers are the basis of many of the
tools with which reasoning is automated. Their common shape is a search
for a countermodel. To check that $\varphi$ is valid, suppose it false
and decompose: the main connective of each formula determines a rule,
branching where the connective demands a choice, closing a branch when it
becomes contradictory.
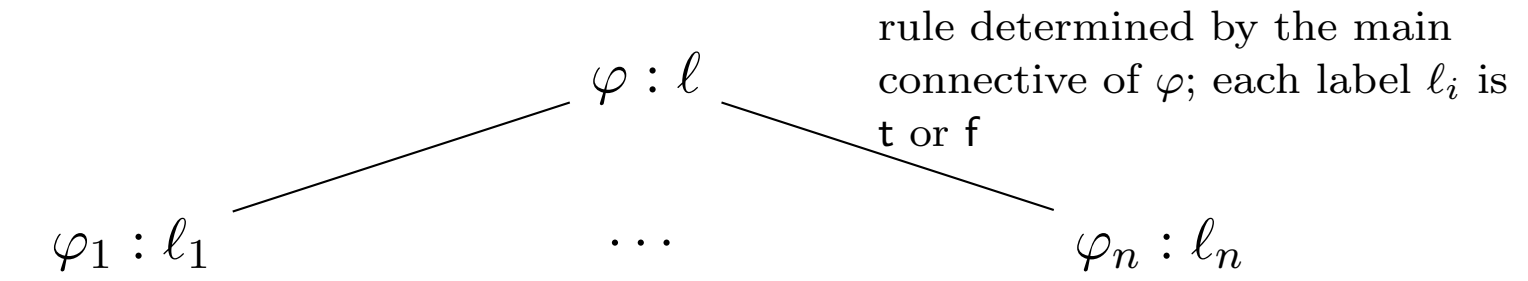
\begin{center}
\begin{tikzpicture}[level distance=11mm, sibling distance=34mm]
\node (root) {$\varphi : \ell$}
  child { node {$\varphi_1 : \ell_1$} }
  child { node {$\cdots$} edge from parent[draw=none] }
  child { node {$\varphi_n : \ell_n$} };
\node[right=9mm of root, font=\scriptsize, text width=42mm]
  {rule determined by the main connective of $\varphi$; each label
   $\ell_i$ is $\mathsf{t}$ or $\mathsf{f}$};
\end{tikzpicture}
\end{center}
Every node carries a label $\ell$, $\ell_1$, \ldots, $\ell_n$ dictating its status in the model under construction. 

Behind these tools stands \emph{model-theoretic semantics} (M-tS): fix a
class of structures, define satisfaction by recursion on the connectives,
and take validity to be truth in every structure. The solvers are how we often use the semantics; a tableau explores
precisely the space of structures the semantics quantifies over. They are
not accessories to the semantics but operationalize it. They are what makes it useful for
representing and automating reasoning --- for \emph{good old-fashioned
AI}.

An alternative semantic paradigm has recently achieved mathematical maturity:
\emph{proof-theoretic semantics} (P-tS)~\cite{SEP-PtS}. Philosophically, this approach differs from M-tS by being based on the
\emph{meaning-as-use} paradigm (as opposed to meaning-as-denotation) imported into logic, with use read as inferential role. Thus the word ``proof''
in the name should not mislead: P-tS is not proof theory, but takes inferences (rather than denotations) as primitive. The philosophical literature here is rich, but it is not our concern in this paper.

We shall concern ourselves with a particularly successful branch of P-tS known as \emph{base-extension semantics} (B-eS).  Here meaning is defined relative to a \emph{base}: a
set of inferential commitments governing atomic sentences. The meaning of a
formula $\varphi$ relative to a base $\Bas$ is given by a judgement
$\suppB{\Bas}\varphi$, called \emph{support} ($\supp$), whose clauses fix, for each
connective, what it takes to support a formula built with it. A base may be thought of as the
P-tS counterpart of a model in M-tS, and support as the counterpart of satisfaction.

The success of B-eS in recent years is its expressiveness. It has now been given for classical propositional
logic~\cite{Sandqvist2009CL} and for intuitionistic propositional logic
(\IPL)~\cite{Sandqvist2015IL}; the relationship between the two is more
delicate than one might expect, and we take it up in Section~\ref{sec:bes}. The
paradigm has since been extended to first-order logic~\cite{FOL},
intuitionistic multiplicative linear logic~\cite{IMLL}, the logic of bunched
implications~\cite{BI}, and modal logics~\cite{Eckhardt}. Given that it has been developed for so many standard logics, the semantics should no
longer be thought exotic. How does one use it though? 
It remains to operationalize the semantics. 

So what stands to B-eS as tableaux and SAT solvers stand to M-tS?  Transposing the
model-search apparatus directly to B-eS would miss exactly the difference that makes
the semantics interesting. The underlying idea of B-eS is behaviour, not structure. Therefore, it seems reasonable to suggest that the  operational counterpart should be a search
through inferential commitments (as opposed to models). Fortunately, there is already a mature technology
for such search, itself a piece of good old-fashioned AI: \emph{logic programming}
(LP). 

Our claim is not that LP may be used to implement B-eS, in the weak sense in
which any mathematics can be implemented. Rather it is that LP operationalizes the unfolding of the semantics as tableaux operationalizes the unfolding of M-tS.   When LP is represented as goal-directed proof-search, it decomposes a \emph{goal} analogously to how tableaux decomposes a formula: the main connective of the goal determines which rule
applies, disjunctive goals branch, and a branch closes successfully when the
goal $G$ has been reduced to something the program $P$ supplies outright: 
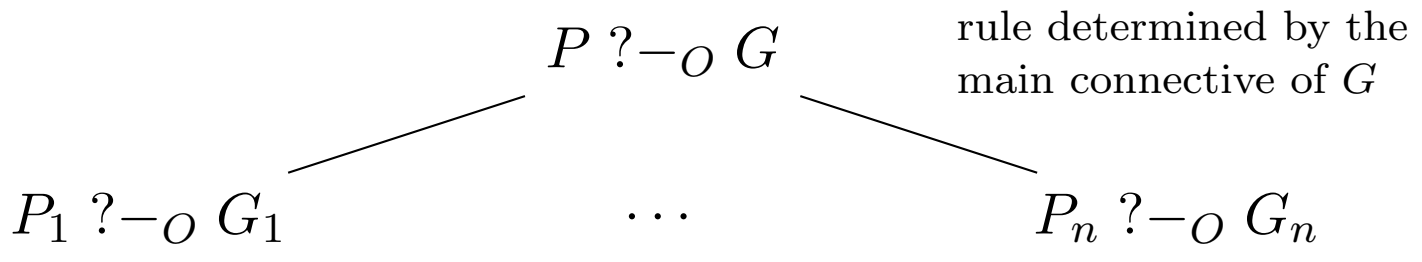
\begin{center}
\begin{tikzpicture}[level distance=11mm, sibling distance=34mm]
\node (root) {$P \opq G$}
  child { node {$P_1 \opq G_1$} }
  child { node {$\cdots$} edge from parent[draw=none] }
  child { node {$P_n \opq G_n$} };
\node[right=9mm of root, font=\scriptsize, text width=42mm]
  {rule determined by the\\ main connective of $G$};
\end{tikzpicture}
\end{center}
We write $P \opder G$ for when the search $P \opq G$ succeeds.

Since B-eS exists for many logics, the claim that it \emph{is} LP has content only once the
right LP language has been matched to the right logic. Here we lean on the
analysis of Miller et al.~\cite{miller1991uniform}: a logic programming
language is given by a class of \emph{definite} formulae, which serve as
program clauses, together with a class of \emph{goal} formulae, which are
queried against them, such that provability is captured by \emph{uniform}
proofs --- goal-directed search in which the logical structure of the goal
drives the computation and the program is consulted only at atoms. Horn
clauses meet the requirement, and are the foundation of Prolog; so do
hereditary Harrop formulae, the foundation of
$\lambda$Prolog~\cite{miller2012programming}.

In this short paper, we show that the B-eS for \IPL{} is captured by a second-order
fragment of hereditary Harrop logic programming, in the family of
$\lambda$Prolog but with quantification over atoms only. That is,
\[
  \suppB{\Bas} \varphi \qquad \text{iff} \qquad \Bas \opder \enc{\varphi},
\]
where $\opder$ is operational derivability and $\enc{\varphi}$ is the encoding
of $\varphi$ as a goal formula. A formula's semantic status can be established
by running it.

The correspondence pays in both directions. Read from right to left, it says
that goal-directed search is not merely a proof-search discipline that happens
to be complete for a well-chosen fragment: it is a semantics. A hereditary
Harrop program is a meaning-conferring object rather than a database of facts,
and the uniform-proof restriction is a semantic condition rather than an
operational convenience.

\medskip
\noindent \textbf{Roadmap.}
Section~\ref{sec:lp} presents the three LP languages we require --- Horn
clauses, hereditary Harrop formulae, and their second-order
extensions --- with their operational semantics and a worked example.
Section~\ref{sec:bes} defines support, explains its clauses, and recalls
the classical/intuitionistic correspondence. Section~\ref{sec:internalize}
develops the encoding and proves the theorem above. Section~\ref{sec:consequences} draws out the consequences and the programme
they suggest. This is a short paper: apart from the cited theorems of
Section~\ref{sec:bes}, the definitions and results of
Sections~\ref{sec:lp}--\ref{sec:internalize} are those of the extended
version~\cite{gheorghiu2026supportsearch}, where full proofs are given; we
indicate the shape of the main argument as we go.

\section{Logic Programming}
\label{sec:lp}
We present LP before B-eS, since some of the machinery can then be
recycled. Throughout we use a proof-theoretic presentation rather than the
resolution-based approach of Kowalski~\cite{kowalski1979logic}. The reason is flexibility:
we shall consider three languages in order of increasing expressiveness. A
language $\mathcal{L}$ comprises a class of \emph{definite formulae} (or
\emph{clauses}) $D$ and a class of \emph{goal formulae} $G$.

\begin{definition}[Program]
\label{def:program}
Let $\mathcal{L}$ be a logic programming language. An $\mathcal{L}$-\emph{program} $P$ is a finite set of definite formulae in the sense of $\mathcal{L}$.
\end{definition}

In the proof-theoretic presentation, \emph{resolution} is replaced by
\emph{backchaining}: to establish an atom, one looks in the program for a
clause whose head is that atom. This is a change of presentation only, not
of substance. A Prolog clause $H :- B_1,...,B_n$ is thus captured by a
definite clause $H \subset B_1,...,B_n$. Here $\subset$ is Kowalski's
reverse implication, read ``$H$ if $B_1,\ldots,B_n$''; in the grammars
below we write clauses head-last, so the same clause appears as
$(B_1 \cap \cdots \cap B_n) \supset H$.

\begin{definition}[Horn clauses]
\label{def:horn}
The language of \emph{propositional Horn clauses} ($\Horn$) consists of the
\emph{definite formulae} $D$ and \emph{goal formulae} $G$ generated by
\[
  D \;::=\; A \;\mid\; G \supset A \;\mid\; D \cap D
  \qquad\qquad
  G \;::=\; A \;\mid\; G \cap G \;\mid\; G \cup G.
\]
\end{definition}

Observe that an implication may occur only in definite formulae, never in
goals. The language can therefore express conditionals but not
hypotheticals --- and hypotheticals are precisely what intuitionistic logic
requires. This motivates an extension in which hypothetical goals are
admitted. Crucially, the extension must be such that proof-search remains
goal-directed, since goal-directedness is the point of the LP formalism.
The following extension is due to Miller~\cite{miller1989logical} and
Harland~\cite{harland1991hereditary}.

\begin{definition}[Hereditary Harrop formulae]
\label{def:hh}
The language of \emph{propositional hereditary Harrop formulae}
(\hH) is generated by
\[
  D \;::=\; A \;\mid\; G \supset A \;\mid\; D \cap D
  \qquad\qquad
  G \;::=\; A \;\mid\; D \supset G \;\mid\; G \cap G \;\mid\; G \cup G.
\]
\end{definition}

Clause heads are atomic, as in the extended
version~\cite{gheorghiu2026supportsearch}: the definite formulae are those of
$\Horn$, and the addition is in the goals. It is the hypothetical goal
$D \supset G$, whose procedural reading is: \emph{to establish $G$ under
the assumption $D$, add $D$ to the program and carry on}. The program is
thus no longer fixed for the duration of a computation; it grows as the
search descends into implicational goals. This is the LP counterpart of
discharging an assumption in natural deduction, and it is the device that
will give implication its intuitionistic content.

For our systems to be genuinely operational, programs must be computable in
some sense. Of course, we could simply require that they have computable
descriptions --- that they be recursively enumerable. However, with one
small addition we can constrain the programs used throughout the paper to
be \emph{finite}, and hence certainly tractable for implementation. The
device, following Miller et al.~\cite{miller1991uniform}, is to observe
that every program we require has a regular structure that can be captured
by second-order quantification. A quantified formula packs a schematic
family --- one instance for each atom --- into a single finite formula.

\begin{definition}[Second-order hereditary Harrop formulae]
\label{def:hh2}
The language $\mathcal{L}^2$ extends $\mathcal{L} \in \{\Horn, \hH\}$ with a \emph{second-order quantifier}
ranging over atoms:
\[
  D \;::=\; \ldots \;\mid\; \forall X\,D
  \qquad\qquad
  G \;::=\; \ldots \;\mid\; \forall X\,G.
\]
\end{definition}

\noindent Here $X$ ranges over a set of propositional variables disjoint
from the atoms. The \emph{atomic formulae} are the atoms together with the
variables, and $A$ ranges over them in the grammars above, so that, for
instance, $G \supset X$ is a definite formula. Programs and goals are
\emph{closed}: every variable occurring in them is bound. We write $G[A/X]$
for the result of replacing the free occurrences of $X$ in $G$ by the atom
$A$, and say that an atom is \emph{fresh} for a formula or program if it
does not occur in it.

We have glossed the procedure as backchaining, which requires looking
inside the program for clauses. Since programs in the above languages have
internal structure, this inspection itself needs a procedural account. The
following provides one.

\begin{definition}[Elaboration]
\label{def:elab}
The elaboration $[P]$ of a program $P$ is defined recursively as follows:
\begin{itemize}
\item $P \subseteq [P]$;
  \item if $D_1 \cap D_2 \in [P]$, then $D_1 \in [P]$ and $D_2 \in [P]$;
  \item if $\forall X\,D \in [P]$, then $D[B/X] \in [P]$ for every atom $B$.
\end{itemize}
\end{definition}

Elaboration acts on clauses one at a time, so $[P]$ is the union
of the elaborations of the members of $P$, and in particular $[P \cup Q] =
[P] \cup [Q]$; we use this freely. Note also that $[P]$ may be infinite,
since the quantifier condition instantiates a quantified clause at every
atom. This is harmless: the operational semantics inspects only finitely
much of it in any one derivation. The important feature is that every
clause eventually elaborates to atoms and to clauses $G \supset A$ with
atomic head.

We may now say what it is to run a query. The relation $\opder$ is defined
uniformly for all three languages, and we give the procedural reading of
each clause alongside it.

\begin{definition}[Operational semantics]
\label{def:op}
The relation $P \opder G$ is defined by:
\begin{itemize}
  \item \textsc{Fact}: $P \opder A$ if $A \in [P]$.
  \item \textsc{Backchain}: $P \opder A$ if $G \supset A \in [P]$ and $P
    \opder G$.
  \item \textsc{And}: $P \opder G_1 \cap G_2$ if $P \opder G_1$ and $P
    \opder G_2$.
  \item \textsc{Or}: $P \opder G_1 \cup G_2$ if $P \opder G_1$ or $P \opder
    G_2$.
  \item \textsc{Augment}: $P \opder D \supset G$ if $P \cup \{D\} \opder G$.
  \item \textsc{Generic}: $P \opder \forall X\,G$ if $P \opder G[A/X]$ for
    some atom $A$ not occurring in $P$ or $G$.
\end{itemize}
\end{definition}

Read as instructions: to establish an atom, find it in the program, or find
a clause with that head and establish the clause's body instead; to
establish a conjunction, establish both parts; to establish a disjunction,
choose a part and establish it; to establish an implication, assume the
antecedent and continue; to establish a schematic claim, establish one
instance at an atom about which nothing is known. Only the first two rules
look at the program at all. The rest are determined by the goal.

\begin{example}
\label{ex:op}
Let $P = \{p \supset q\}$. The following query succeeds; each edge of the
search tree is labelled with the rule applied.
\begin{center}
\begin{tikzpicture}[redtree]
\node (n0) {$P \opq \forall X\,((q \supset X) \supset (p \supset X))$};
\node (n1) [below=of n0] {$P \opq (q \supset A) \supset (p \supset A)$};
\node (n2) [below=of n1] {$P, q \supset A \opq p \supset A$};
\node (n3) [below=of n2] {$P, q \supset A, p \opq A$};
\node (n4) [below=of n3] {$P, q \supset A, p \opq q$};
\node (n5) [below=of n4] {$P, q \supset A, p \opq p$};
\node (n6) [below=of n5] {\checkmark};
\draw[->] (n0) -- (n1) node[rulelab] {\textsc{Generic}, $A$ fresh};
\draw[->] (n1) -- (n2) node[rulelab] {\textsc{Augment}};
\draw[->] (n2) -- (n3) node[rulelab] {\textsc{Augment}};
\draw[->] (n3) -- (n4) node[rulelab] {\textsc{Backchain} on $q \supset A$};
\draw[->] (n4) -- (n5) node[rulelab] {\textsc{Backchain} on $p \supset q$};
\draw[->] (n5) -- (n6) node[rulelab] {\textsc{Fact}};
\end{tikzpicture}
\end{center}
The goal expresses, schematically, that whatever follows from $q$ follows
from $p$. The fresh atom $A$ stands proxy for an arbitrary conclusion, and
the two applications of \textsc{Augment} show the program growing as the
search descends: what began as a single clause is, by the fourth line, a
program of three.
\end{example}

The freshness condition in \textsc{Generic} is important. A
schematic claim would seem to call for infinitely many checks, and the rule
performs only one. It does not require that $G[A/X]$ hold for every atom; it
requires it for some atom occurring in neither the program nor the goal.
Because nothing whatever is known about such an atom, a derivation using it
can appeal to no property it does not share with every other atom, and so
the single witness establishes the schema. This is the eigenvariable
discipline of proof theory, and it is what will make the universal
quantifiers of the support semantics finitary --- a point on which the main
theorem depends.

It remains to record three structural properties of $\opder$, each of which
says something one would want to be true of a search procedure. We write $Q
\baseext P$ to mean $Q \supseteq P$, borrowing the notation for base
extension deliberately: Section~\ref{sec:internalize} will identify the two
relations. All three properties are proved by routine induction on
derivations, and all three preserve the height of the derivation --- which
matters, since the induction in Section~\ref{sec:internalize} is on height.

\begin{lemma}[Monotonicity]
\label{lem:mono}
If $P \opder G$ and $Q \baseext P$, then $Q \opder G$.
\end{lemma}

\begin{lemma}[Renaming]
\label{lem:rename}
Let $\sigma$ be any function from atoms to atoms, extended homomorphically
to formulae and programs. If $P \opder G$, then $P\sigma \opder G\sigma$.
\end{lemma}

\begin{lemma}[Inversion]
\label{lem:inv}
If $P \opder D \supset G$, then $P \cup \{D\} \opder G$. If $P \opder G_1
\cap G_2$, then $P \opder G_i$ for each $i$. If $P \opder \forall X\,G$,
then $P \opder G[A/X]$ for every atom $A$.
\end{lemma}

Monotonicity says that adding clauses never destroys a solution. Renaming
says that the names of atoms carry no information, and requires care only at
\textsc{Generic}, where the eigenvariable must be moved out of the way of
$\sigma$ before the induction hypothesis applies; Monotonicity uses it at the
same point, since the eigenvariable of a search over $P$ may occur in the
larger program. Inversion is the converse
of the goal-directedness we have been emphasising: because the main
connective of a goal determines the last rule of its derivation ---
\textsc{Fact} and \textsc{Backchain} concluding only atomic goals --- one
can read the premisses back off the conclusion; for the quantifier, Renaming
then moves the eigenvariable of \textsc{Generic} to any atom whatever. It is the rule-by-rule
character of $\opder$ that makes this possible, and it is what will let the
comparison with support in Section~\ref{sec:internalize} run in both
directions.

\section{Base-extension Semantics}
\label{sec:bes}
We have already given a brief introduction to B-eS above, and we elide
further motivation in favour of the technical details. Our logical syntax
is given by the grammar
\[
  \varphi, \psi \;::=\; A \;\mid\; \bot \;\mid\;
    \varphi \land \psi \;\mid\; \varphi \lor \psi \;\mid\; \varphi \to \psi,
\]
with $\lnot\varphi \coloneqq \varphi \to \bot$. Throughout, let $\mathcal{L}$ be one of the LP languages of Section~\ref{sec:lp}.

\begin{definition}[Base]
\label{def:base}
 An
\emph{$\mathcal{L}$-base} $\Bas$ is an $\mathcal{L}$-program. We
write $\BasC \baseext \Bas$ for extension of bases, meaning $\BasC \supseteq
\Bas$ with $\BasC$ an $\mathcal{L}$-base.
\end{definition}

It may seem redundant to define bases and programs separately; the point is
to smooth the correspondence between the two literatures. Nothing in the
literature on B-eS requires bases to be finite --- indeed, it is crucial
that they may be infinite. The restriction to finite bases is available
here only because of the second-order quantification: readers within the
P-tS literature will recognise that this recovers the expressive strength
required for the results cited below.

The support relation is defined relative to a fixed choice of
$\mathcal{L}$; the base extensions quantified over in its clauses range
over $\mathcal{L}$-bases. Where the choice needs emphasis we speak of
$\mathcal{L}$-support.

\begin{definition}[Support]
\label{def:support}
The relations $\suppB{\Bas} \varphi$ and $\Gamma \suppB{\Bas} \varphi$ are
defined by the clauses of Figure~\ref{fig:support}, where $\Gamma$ is a
finite set of formulae and we write $\varphi \suppB{\Bas} \psi$ for
$\{\varphi\} \suppB{\Bas} \psi$.
\end{definition}

\begin{figure}[t]
\hrule\vspace{2mm}
\begin{align*}
  &\supp_{\Bas} A
    &&\;\text{ iff }\; \Bas \opder A \tag{At} \\[1mm]
  &\supp_{\Bas} \varphi \to \psi
    &&\;\text{ iff }\; \varphi \supp_{\Bas} \psi \tag{$\to$} \\[1mm]
  &\supp_{\Bas} \varphi \land \psi
    &&\;\text{ iff }\;
      \text{$\supp_{\Bas} \varphi$ and $\supp_{\Bas} \psi$}
      \tag{$\land$} \\[1mm]
  &\supp_{\Bas} \varphi \lor \psi
    &&\;\text{ iff }\;
      \text{for every $\BasC \baseext \Bas$ and $A \in \At$,}
      \notag\\
    & &&\qquad\quad  \text{if $\varphi \supp_{\BasC} A$ and $\psi \supp_{\BasC}
      A$, then $\supp_{\BasC} A$} \tag{$\lor$} \\[1mm]
  &\supp_{\Bas} \bot
    &&\;\text{ iff }\; \supp_{\Bas} A \text{ for every } A \in \At
    \tag{$\bot$} \\[1mm]
  &\Gamma \supp_{\Bas} \varphi
    &&\;\text{ iff }\;
      \text{for every $\BasC \baseext \Bas$,
      if $\supp_{\BasC} \psi$ for all $\psi \in \Gamma$,
      then $\supp_{\BasC} \varphi$} \tag{Inf}
\end{align*}
\vspace{1mm}\hrule
\caption{Support in a base.}
\label{fig:support}
\end{figure}

The clauses ($\lor$) and ($\bot$) give the meaning of a connective as a
consequence condition: they state what one may \emph{conclude} from the
formula, an atom $A$ standing proxy for an arbitrary conclusion. The formula
$\varphi \lor \psi$ licenses every conclusion that each disjunct
separately licenses, and the formula $\bot$ licenses every conclusion. Each is, visibly, the
elimination rule for its connective read as a definition of meaning. The
clause ($\to$) fits the same pattern --- $\varphi \to \psi$ licenses $\psi$
from $\varphi$ --- expressed through the context-extension clause (Inf).
The clause ($\land$) is direct, but it too has an elimination form
(Remark~\ref{rem:conj}). The clause (At)
is the interface with the base: atomic support is provability in the base,
and here the reading of bases as programs does its first work, since
$\opder$ is proof-search.

\begin{remark}
\label{rem:conj}
The clauses of Figure~\ref{fig:support} are Sandqvist's~\cite{Sandqvist2015IL}.
The direct clause for conjunction is equivalent to an elimination-style
clause of the same shape as ($\lor$)~\cite{IMLL}:
\[
\supp_\Bas \varphi \land \psi \qquad \mbox{iff} \qquad
\text{for every $\BasC \baseext \Bas$ and $A \in \At$,
      if $\varphi, \psi \supp_{\BasC} A$ then $\supp_{\BasC} A$},
\]
and implication likewise admits an elimination form: $\supp_\Bas \varphi
\to \psi$ iff for every $\BasC \baseext \Bas$ and $A \in \At$, if
$\supp_\BasC \varphi$ and $\psi \supp_\BasC A$ then $\supp_\BasC A$
\cite[Corollary~3.7]{gheorghiu2026supportsearch}. Every connective is thus
characterised by what it licenses one to conclude.
\end{remark}

\begin{example}
\label{ex:supp-or}
Fix $\mathcal{L} = \Horn$.
Let $\Bas = \{p\}$ and consider $\supp_\Bas p \lor q$. Does it hold? It
does. Given $\BasC \baseext \Bas$ and an atom $A$ with $p \supp_\BasC A$
and $q \supp_\BasC A$, the fact $p \in \BasC$ yields $\supp_\BasC p$ by
(At), and the first hypothesis then yields $\supp_\BasC A$. Note that no
disjunction occurs in the base: the disjunction is supported through what
its disjuncts license, not through any disjunctive commitment.
\end{example}

A formula $\varphi$ is \emph{$\mathcal{L}$-valid} iff $\supp_\Bas \varphi$
holds for every $\mathcal{L}$-base $\Bas$. Equivalently,  $\varphi$ is \emph{$\mathcal{L}$-valid} iff $\supp_{\emptyset} \varphi$.
What is curious about B-eS is
that the same clauses define two different logics depending on the notion of base:

\begin{theorem}[Sandqvist~\cite{Sandqvist2009CL}]
\label{thm:classical}
A formula $\varphi$ is $\Horn^2$-valid if and only if $\varphi$ is provable
in classical propositional logic.
\end{theorem}

\begin{theorem}[Sandqvist~\cite{Sandqvist2015IL}]
\label{thm:intuitionistic}
A formula $\varphi$ is $\hH^2$-valid if and only if $\varphi$ is provable
in intuitionistic propositional logic.
\end{theorem}

\noindent Sandqvist states Theorem~\ref{thm:intuitionistic} for his class of
bases, arbitrary sets of atomic rules of level at most two; the proof
transfers unchanged to $\hH^2$-bases~\cite[Section~3]{gheorghiu2026supportsearch}.

There is currently no consensus on how to understand this result. The
contravariance --- $\hH$ extends $\Horn$, while classical logic extends
intuitionistic logic --- makes some sense: in $\hH$ the support clauses
quantify over a larger space of bases than in $\Horn$, so validity is
harder to attain and fewer formulae achieve it. But this heuristic does not
account for why one should expect exactly classical logic in the one case
and exactly intuitionistic logic in the other. The key differentiating
feature appears to be the hypothetical reasoning available in $\hH$ but not
in $\Horn$.

\begin{example}[Peirce's formula]
\label{ex:peirce}
Peirce's formula $((p \to q) \to p) \to p$ is classically valid, hence
$\Horn^2$-valid by Theorem~\ref{thm:classical}; but it is not provable in
\IPL, hence not $\hH^2$-valid by Theorem~\ref{thm:intuitionistic}. So there
is a hereditary Harrop base --- we shall see in
Example~\ref{ex:peirce-run} that the empty base suffices --- at which the
support clauses refuse it. The refusal is possible only because
\hH-extensions can install the hypothesis $(p \to q) \to p$ as a genuinely
hypothetical commitment rather than a material one.
\end{example}

This completes our introduction to B-eS. The remainder of the paper
establishes that LP, as defined above, operationalises it.

\section{Soundness and Completeness}
\label{sec:internalize}

We have defined both LP and B-eS; it remains to explain their correspondence.
The idea is that support judgements correspond to certain queries. Given the
complexity of the clauses in Figure~\ref{fig:support}, we restrict attention to
$\hH^2$, which is expressive enough to encapsulate all the requisite structure.
By Theorem~\ref{thm:intuitionistic}, this means that in this section we are
concerned only with $\IPL{}$; classical logic is treated in
Section~\ref{sec:consequences}.

Every LP language we consider has two syntactic classes of formula: \emph{goals},
which are queried, and \emph{definite clauses}, which make up the program.
Roughly, these play the roles of the query and the clause database in a
resolution setting, but with one crucial difference: the rule \textsc{Augment},
on encountering a goal $D \supset G$, installs $D$ into the program as a
hypothesis and continues with $G$. Consequently an encoded formula may appear in
either class --- as a goal when it is queried, and as a program clause when
\textsc{Augment} installs it. Since $\IPL{}$ has only one class of formulae, we
work with the fragment of $\hH^2$ that makes no such distinction.

\begin{definition}[Neutral formulae]
\label{def:neutral}
A formula is \emph{neutral} if it is generated by the grammar
\[
  N \;::=\; A \;\mid\; N \supset A \;\mid\; N \cap N \;\mid\; \forall X\,N,
\]
where $A$ ranges over atomic formulae.
\end{definition}

\begin{proposition}
\label{prop:neutral}
Every neutral formula is both a definite formula and a goal formula of $\hH^2$.
\end{proposition}

\noindent
The proof is a routine induction on the grammar: an implication $N \supset A$
is definite read as $G \supset A$ and a goal read as $D \supset G$, and the
remaining cases are immediate. Note that the consequent of an implication in a
neutral formula is atomic: a $\cup$-free implication with a compound
consequent is a goal but not a definite formula.

We now encode $\IPL{}$ formulae as neutral formulae.

\begin{definition}[Encoding]
\label{def:encoding}
To each $\IPL{}$ formula $\varphi$, assign a formula $\enc{\varphi}$ by:
\[
  \begin{aligned}
    \enc{A} &\;=\; A \\[3pt]
    \enc{\bot} &\;=\; \forall X\, X \\[3pt]
    \enc{\varphi \to \psi} &\;=\; \forall X\,
      \bigl((\enc{\varphi} \cap (\enc{\psi} \supset X)) \supset X \bigr) \\[3pt]
    \enc{\varphi \land \psi} &\;=\; \enc{\varphi} \cap \enc{\psi} \\[3pt]
    \enc{\varphi \lor \psi} &\;=\; \forall X\,
      \bigl(((\enc{\varphi} \supset X) \cap (\enc{\psi} \supset X))
      \supset X \bigr)
  \end{aligned}
\]
where $X$ is in each case a variable not occurring in $\enc{\varphi}$ or
$\enc{\psi}$.
\end{definition}

By induction on $\varphi$, each $\enc{\varphi}$ is neutral, so a base
extended by an encoded formula is again a base. The encoding transcribes
the support clauses. The goal $\enc{\varphi \lor \psi}$ says: for each atom
$X$ taken as an arbitrary conclusion, if $\enc{\varphi}$ yields $X$ and
$\enc{\psi}$ yields $X$, then $X$ --- this is clause ($\lor$), with the
universal over atoms become $\forall X$ and the quantifier over base
extensions become program augmentation. The goal $\enc{\bot} = \forall X\,X$
says that every conclusion follows. Conjunction is encoded directly,
following clause ($\land$). Implication is encoded in elimination form: the
goal $\enc{\varphi \to \psi}$ says that, for an arbitrary conclusion $X$,
$\enc{\varphi}$ together with the continuation $\enc{\psi} \supset X$
yields $X$ --- the elimination form of ($\to$) recorded in
Remark~\ref{rem:conj}. The direct encoding $\enc{\varphi} \supset
\enc{\psi}$ is the natural first thought, but its consequent is in general
compound, so it is not a definite formula and cannot be installed in a base
by \textsc{Augment}; the elimination form keeps every clause head
atomic~\cite{gheorghiu2026supportsearch}. The image of the encoding uses
only atoms, $\supset$, $\cap$, and $\forall$: the goal-only connective
$\cup$ never occurs.

\begin{remark}[Nomenclature]
\label{rem:rp}
Definitions of the connectives from $\supset$ and a propositional
quantifier go back to Russell, and their proof-theoretic study to Prawitz;
the encoding is accordingly the \emph{Russell--Prawitz translation} of the
connectives~\cite{russell1903,Prawitz2006natural}, here with the quantifier
restricted to atomic instantiation as in the atomic polymorphism literature~\cite{Ferreira2006,Ferreria2013}
(see also Remark~\ref{rem:nosubst}). Functional programmers will recognise the
shape of a continuation-passing transform --- the atom $X$ is a
continuation, an arbitrary conclusion to be delivered --- and we retain
that reading informally, but we do not call the encoding CPS: the term has
a precise, and different, meaning as a transformation on source
programs.
\end{remark}

\begin{remark}
    The encoding $\enc{\cdot}$ is not one clever translation among many. It is
determined, connective by connective, by the support semantics:
conjunction is transcribed from its clause directly, and $\bot$, $\lor$,
and $\to$ from their elimination behaviour, by replacing the quantifier
over atoms with $\forall$ and the quantifier over base extensions with
$\supset$. This canonicity suggests an explanation of the coincidence observed
in~\cite{MINTS} between Mints' clausal translation of \IPL{} and
Sandqvist's completeness construction: both are arrived at by writing the
elimination rules as clauses over atomic vocabularies --- instances of the
same Russell--Prawitz transcription, reached from different directions.
\end{remark}

Before proving the encoding faithful, it is worth asking how neutral formulae sit
within LP. Atoms are the paradigm of a formula that is simultaneously a goal and
a definite clause, and neutral formulae extend that class; so one might expect
them to behave like atoms in every respect --- in particular, to be substitutable
for them. They are not, and the reason is instructive.

In full second-order propositional logic a universal quantifier may be
instantiated at an arbitrary formula. The quantifier of $\hH^2$ is not of this
kind: operationally it is \emph{atomic}. \textsc{Generic} discharges a universal
goal by introducing a single fresh atom, and elaboration instantiates a universal
clause only at atoms. This restriction is what keeps proof search effective, and
it is exactly what separates neutral formulae from atoms.

\begin{remark}[Failure of compound instantiation]
\label{rem:nosubst}
It is tempting to posit the following substitution principle: if $P \opder G$ and
the atom $p$ does not occur in $P$, then $P \opder G[F/p]$ for every neutral $F$.
It fails. Take
\[
  P = \{\, \forall Y\,(Y \supset q) \,\}, \qquad G = p \supset q,
\]
where $p$, $q$, and $a$ are distinct atoms.

\emph{The atomic instance succeeds.} \textsc{Augment} turns $P \opder p \supset q$
into the query $P, p \opq q$. As $p$ is an atom, $p \supset q$ is an instance of
the clause of $P$ and so lies in the elaboration $[P]$; \textsc{Backchain} on it
leaves $P, p \opq p$, which \textsc{Fact} closes.

\emph{The compound instance fails.} Substituting $F = a \supset a$ for $p$ yields
the goal $(a \supset a) \supset q$, and after \textsc{Augment} the query
$P, a \supset a \opq q$. Since the clause of $P$ elaborates at atoms only, the
clauses of $[P \cup \{a \supset a\}]$ with head $q$ are precisely those of the
form $B \supset q$ with $B$ atomic, so \textsc{Backchain} demands that some atom
be derivable. None is: the only heads available are $q$, which itself requires an
atom, and $a$, which requires $a$. Hence
$P \not\opder (a \supset a) \supset q$.

\emph{Atomic polymorphism.} 
By \textsc{Generic}, the successful half says precisely that
$P \opder \forall X\,(X \supset q)$. So the universal goal is derivable while one
of its compound instances is not: the quantifier of $\hH^2$ is genuinely atomic.
This is the operational face of the predicativity of atomic
polymorphism~\cite{Ferreira2006}, and it carries a methodological consequence.
Theorem~\ref{thm:main} cannot be proved by a substitution lemma. The arguments
below use instead renaming (Lemma~\ref{lem:rename}), which replaces atoms by
atoms, together with identity and cut for neutral formulae.
\end{remark}

Nonetheless, some of the behaviour of atoms does survive, and it is exactly the
behaviour proof search relies on. The rule \textsc{Fact}, which closes an atomic
goal sitting in the elaborated program, generalises to neutral formulae; so does
the admissibility of cut.

\begin{lemma}[Identity]
\label{lem:id}
Let $F$ be neutral. If $F \in [P]$, then $P \opder F$. In particular,
$P, F \opder F$ for every program $P$.
\end{lemma}

\begin{lemma}[Cut]
\label{lem:cut}
Let $F$ be neutral. If $P \opder F$ and $Q, F \opder G$, then $P, Q \opder G$.
\end{lemma}

\noindent
Note that neither statement would even typecheck without
Proposition~\ref{prop:neutral}: in each, $F$ occurs once as a goal and once as a
program clause. This is where membership in both syntactic classes earns its
keep.

Identity is proved by induction on the size of $F$, where $F[A/X]$ has the
same size as $F$. Cut is proved by induction on the size of $F$ and, within
it, on the height of the derivation of $Q, F \opder G$. Its key step is
that every member of $[F]$ is neutral, of size at most that of $F$, and
derivable from $P$ whenever $F$ is, by inversion
(Lemma~\ref{lem:inv}); renaming (Lemma~\ref{lem:rename}) handles the
eigenvariable introduced by \textsc{Generic}. The arguments are those of
cut-admissibility for uniform proofs~\cite{miller1991uniform}, adapted to
the atomic quantifier. Details are in the extended
version~\cite{gheorghiu2026supportsearch}.

We can now return to the picture from the introduction. A tableau decomposes a
formula according to its main connective; goal-directed search decomposes a goal
in the same way. The encoding of Definition~\ref{def:encoding} makes the two
decompositions line up with a third: the clauses of Figure~\ref{fig:support}.
Each connective of \IPL{} is sent to a neutral formula whose outermost structure
determines which rule of $\hH^2$ fires, and that rule re-enacts the support
clause for that connective. Conjunction is sent to $\cap$, where \textsc{And}
splits the goal. Falsum is sent to $\forall X\,X$, where \textsc{Generic} introduces a fresh
atom to be derived. Implication and disjunction are sent to universal
formulae, where \textsc{Generic} introduces a fresh atom, \textsc{Augment}
installs the continuation clauses just as the support clauses quantify over
base extensions, and \textsc{Backchain} consumes them.

\begin{theorem}[Soundness and Completeness]
\label{thm:main}
For every base $\Bas$ and \IPL{} formula $\varphi$:
\[
  \suppB{\Bas}\varphi
  \qquad\text{if and only if}\qquad
  \Bas \opder \enc{\varphi}.
\]
\end{theorem}

\begin{proof}[Proof sketch]
By simultaneous induction on $\varphi$, both directions together. The case
of atoms is clause (At). For $\bot$, \textsc{Generic} at an atom fresh for
$\Bas$ gives one direction and inversion the other. For $\land$, clause
($\land$), the induction hypothesis, \textsc{And}, and inversion suffice. For $\lor$ and $\to$, left
to right: \textsc{Generic} and \textsc{Augment} reduce the encoded goal to
proving a fresh atom $A$ from $\BasC = \Bas \cup \{D\}$, where $D$ is the
continuation clause --- $(\enc{\varphi_1} \supset A) \cap (\enc{\varphi_2}
\supset A)$ for $\lor$, and $\enc{\varphi_1} \cap (\enc{\varphi_2} \supset A)$
for $\to$. For $\lor$, the induction hypothesis and \textsc{Backchain} on
$\enc{\varphi_i} \supset A$ give $\varphi_i \suppB{\BasC} A$ for each $i$,
and clause ($\lor$) delivers $\suppB{\BasC} A$. For $\to$, identity and the
induction hypothesis give $\suppB{\BasC} \varphi_1$, clauses ($\to$) and
(Inf) give $\suppB{\BasC} \varphi_2$, and the induction hypothesis and
\textsc{Backchain} on $\enc{\varphi_2} \supset A$ give $\BasC \opder A$.
Right to left: given the extension $\BasC$ and atom $A$ supplied by clause
($\lor$), or by the elimination form of ($\to$) (Remark~\ref{rem:conj}),
monotonicity and inversion unwind \textsc{Generic} and \textsc{Augment} to
give $\BasC, D \opder A$; identity, the induction hypothesis,
\textsc{Augment}, and \textsc{And} give $\BasC \opder D$; and cut gives
$\BasC \opder A$. Since the
induction hypothesis is used in both directions, the two are proved
together. Compound instantiation is never used, in keeping with
Remark~\ref{rem:nosubst}: the fresh atom of \textsc{Generic} is moved to the
atom supplied by the support clause by inversion, which is to say by
renaming. The full proof is~\cite[Theorem~4.2]{gheorghiu2026supportsearch}.
\end{proof}

\begin{corollary}
\label{cor:exec}
$\varphi$ is valid in \IPL{} if and only if
$\emptyset \opder \enc{\varphi}$.
\end{corollary}

\begin{example}
\label{ex:conj-run}
The formula $(p \land q) \to p$ encodes as
$\forall X\,(((p \cap q) \cap (p \supset X)) \supset X)$, since
$\enc{p \land q} = p \cap q$.
The query succeeds from the empty program:
\begin{center}
\begin{tikzpicture}[redtree]
\node (n0) {$\emptyset \opq \enc{(p \land q) \to p}$};
\node (n1) [below=of n0] {$\emptyset \opq ((p \cap q) \cap (p \supset A)) \supset A$};
\node (n2) [below=of n1] {$D_0 \opq A$};
\node (n3) [below=of n2] {$D_0 \opq p$};
\node (n4) [below=of n3] {\checkmark};
\draw[->] (n0) -- (n1) node[rulelab] {\textsc{Generic}, $A$ fresh};
\draw[->] (n1) -- (n2) node[rulelab]
  {\textsc{Augment}, $D_0 = (p \cap q) \cap (p \supset A)$};
\draw[->] (n2) -- (n3) node[rulelab]
  {\textsc{Backchain} on $p \supset A \in [D_0]$};
\draw[->] (n3) -- (n4) node[rulelab] {\textsc{Fact}, $p \in [D_0]$};
\end{tikzpicture}
\end{center}
The fresh atom $A$ is the conclusion to be delivered. Backchaining passes
control to the continuation $p \supset A$, and the conjunct $p$ is then
found in the elaborated program: the elimination rule for conjunction is
executed as a projection.
\end{example}

\begin{example}[Peirce's formula, executed]
\label{ex:peirce-run}
Write
$a = \enc{p \to q} = \forall Y\,((p \cap (q \supset Y)) \supset Y)$ and
$h = \enc{(p \to q) \to p} = \forall Z\,((a \cap (p \supset Z)) \supset Z)$,
so that $\enc{((p \to q) \to p) \to p} = \forall X\,((h \cap (p \supset X))
\supset X)$. The search from the empty program runs as follows:
\begin{center}
\begin{tikzpicture}[redtree]
\node (n0) {$\emptyset \opq \enc{((p \to q) \to p) \to p}$};
\node (n1) [below=of n0] {$D \opq A$};
\node (n2) [below=of n1] {$D \opq a$};
\node (n3) [below=of n2] {$D, p \cap (q \supset B) \opq B$};
\node (n4) [below=of n3] {$D, p \cap (q \supset B) \opq q$};
\node (n5) [below=of n4] {$D, p \cap (q \supset B) \opq a$};
\node (n6) [below=of n5] {$\vdots$};
\draw[->] (n0) -- (n1) node[rulelab]
  {\textsc{Generic}, \textsc{Augment}, $D = h \cap (p \supset A)$};
\draw[->] (n1) -- (n2) node[rulelab]
  {\textsc{Backchain} on $(a \cap (p \supset A)) \supset A \in [D]$, \textsc{And}};
\draw[->] (n2) -- (n3) node[rulelab] {\textsc{Generic}, \textsc{Augment}};
\draw[->] (n3) -- (n4) node[rulelab] {\textsc{Backchain} on $q \supset B$};
\draw[->] (n4) -- (n5) node[rulelab]
  {\textsc{Backchain} on $(a \cap (p \supset q)) \supset q \in [D]$, \textsc{And}};
\end{tikzpicture}
\end{center}
The alternatives lead to the same place. At the second line, \textsc{Backchain}
on $p \supset A$ demands $p$, and the only clauses with head $p$ are instances
of $h$, which demand $a$; at the fourth, \textsc{Backchain} on the instance of
$h$ at $B$ demands $a$ directly. So every branch returns to the goal $a =
\enc{p \to q}$, and establishing $a$ requires $q$, whose only clauses are
instances of $h$ demanding $a$ again. Any derivation would therefore contain
a strictly shorter one of the same kind, so there is none: $\emptyset
\not\opder \enc{((p \to q) \to p) \to p}$, which by Corollary~\ref{cor:exec}
witnesses the intuitionistic invalidity of Peirce's formula --- and, by
Theorem~\ref{thm:main}, the failure of support at the empty base promised
in Example~\ref{ex:peirce}. The formula is classically valid; what the
failed search records is that the hypothesis $(p \to q) \to p$, installed
as a genuinely hypothetical commitment, yields $p$ only given $p \to q$, and
nothing yields $q$.
\end{example}

This operationalizes the B-eS for \IPL{}. We motivated it as the analogue
of tableaux for P-tS, and before closing the section that motivation merits
some final remarks. The analogy breaks in three places, but each break is
the M-tS/P-tS difference showing through.

First, polarity. Every node of a tableau carries a label $\mathsf{t}$ or
$\mathsf{f}$, recording the status the formula is to have in the model under
construction; the search is a refutation, begun by supposing $\varphi$ false. A
query $P \opq G$ carries no such label. There is nothing to suppose false,
because support is not truth-in-a-structure and its negation is not a constraint
one can propagate. The search is direct: it looks for the commitments that would
confer the meaning, not for a structure witnessing that they fail.

Second, and consequently, the two searches read their branches dually. A tableau
succeeds by closing every branch and refutes by leaving one open, the open branch
being the countermodel it has built. Goal-directed search succeeds on a
\emph{single} branch and fails only when no branch can close; and when it fails it
hands back nothing. Example~\ref{ex:peirce-run} produces no object witnessing the
failure of support --- only the observation that every branch returns to the same
unmet demand. In M-tS a
refutation must exhibit a structure, since validity is truth in all of them. Here
the base is given in advance, and the quantification over extensions has been
absorbed into the support clauses themselves. There is no countermodel to build
because there is nothing left to construct.

Third, the program grows. \textsc{Augment} installs a hypothesis and the search
continues against a larger program, so the object consulted at the leaves is not
fixed in advance. A tableau branch also grows, but it grows by decomposition:
what appears is always a subformula of what was there. This is what
Example~\ref{ex:peirce-run} turns on. Peirce's formula fails not because some
branch stays open, but because the hypothesis $(p \to q) \to p$, once
installed as a genuine commitment, licenses nothing that can be discharged.

None of this is a defect of the analogy; it is the analogy doing its work. We
should not expect the operationalization of a use-based semantics to look like
the operationalization of a denotational one, and the respects in which the two
differ are exactly the respects in which B-eS differs from M-tS. What survives is
the structural claim with which we began: a rule determined by the main
connective, branching where the connective demands a choice, and closure when the
decomposition reaches something the ambient object supplies outright. That claim
survives intact, and Corollary~\ref{cor:exec} shows it survives with something to
spare. Validity in M-tS is a quantification over structures, which a tableau
discharges by constructing the one candidate that matters. Validity here is a
single query against the empty program. The quantifier that a tableau must
simulate has already been paid for by the semantics.

\section{Extensions and classical logic}
\label{sec:consequences}

We have established the operationalization only for the B-eS of \IPL{}, but such
semantics now exist for a range of logics: first-order logic~\cite{FOL}, linear
logic~\cite{IMLL}, the logic of bunched implications~\cite{BI}, and modal
logics~\cite{Eckhardt}. In each case the support clauses have the shape that drove
the present construction --- a quantification over base extensions, discharged at
atoms. The programme this paper suggests is accordingly: for each such logic, find
the LP language whose operational steps internalise those quantifiers. The
expected pairings are informative --- first-order B-eS with first-order hereditary
Harrop programs~\cite{miller1991uniform,harland1991hereditary}, IMLL with the linear
hereditary Harrop programs of Lolli~\cite{hodas1994logic} --- and the first-order
case is the natural first test, both sides of the correspondence being already in
place.

In every case the challenge is one of expressiveness, and classical logic shows
what is at stake. Theorem~\ref{thm:classical} identifies B-eS over $\Horn^2$ with
classical logic, so the target language is fixed for us; but the encoding of
Definition~\ref{def:encoding} cannot live there. Its work is done by
$\forall X\,(\cdots \supset X)$ --- a continuation abstracted over, then supplied
--- and while $\Horn^2$ has the quantifier, it has no hypothetical goals, so
neither the goal $\forall X\,(\cdots \supset X)$ nor its body is available there. The pairing that Theorem~\ref{thm:classical} hands us is therefore the one
pairing for which we have no encoding, and the machinery that made \IPL{} work is
exactly what $\Horn^2$ withholds.

Classical \emph{validity}, on the other hand, is already within reach. By
Glivenko's theorem, $\varphi$ is classically valid if and only if
$\neg\neg\varphi$ is intuitionistically valid, so by Corollary~\ref{cor:exec} it
suffices to run
\[
  \emptyset \opq \enc{\neg\neg\varphi}
  \qquad\text{where}\qquad
  \enc{\neg\varphi} = \forall Y\,\bigl((\enc{\varphi} \cap
    ((\forall X\, X) \supset Y)) \supset Y\bigr) .
\]
But this is a reduction, not an
operationalization. It decides classical validity by running an intuitionistic
query, and so tells us nothing about how a classical base confers meaning; the
double negation is a device of the translation rather than anything the search
enacts. It also does not travel: Glivenko's theorem is propositional, so the route
closes precisely where the programme above says to go next.

A native treatment would have to make the classical structure operational rather
than translate it away, and the continuation reading suggests where to look. In
the typed $\lambda$-calculus, continuations at classical type are control
operators~\cite{griffin1990}, which points to an operational semantics with a
control facility --- a restart, or a backtracking-to-continuation rule --- layered
over Horn programs. This would keep the program language that
Theorem~\ref{thm:classical} demands while recovering operationally the expressive
power the encoding obtained from $\forall$. Whether the resulting notion of
success matches support over a Horn base is the obvious question, and we leave it
open.

\bibliographystyle{plain}
\bibliography{bib}

\end{document}